\documentclass[aps,prl,twocolumn,superscriptaddress,floatfix,letter,longbibliography]{revtex4-2}

\usepackage{epsfig,amsmath,amssymb,color,amsfonts,physics,microtype}
\usepackage[bookmarks=true,colorlinks,citecolor=red]{hyperref}
\usepackage{dsfont}
\usepackage{wrapfig}
\usepackage{tikz}
\usepackage{physics}
\usepackage{mathrsfs}
\usepackage[export]{adjustbox}
\usetikzlibrary{quantikz}
\usepackage{soul}
\usepackage{comment}

\begin{document}

\title{Clifford Dressed Time-Dependent Variational Principle}
\author{Antonio Francesco Mello}
\affiliation{International School for Advanced Studies (SISSA), via Bonomea 265, 34136 Trieste, Italy}
\author{Alessandro Santini}
\affiliation{International School for Advanced Studies (SISSA), via Bonomea 265, 34136 Trieste, Italy}
\author{Guglielmo Lami}
\affiliation{Laboratoire de Physique Th\'eorique et Mod\'elisation, CNRS UMR 8089, CY Cergy Paris Universit\'e, 95302 Cergy-Pontoise Cedex, France}
\author{Jacopo De Nardis}
\affiliation{Laboratoire de Physique Th\'eorique et Mod\'elisation, CNRS UMR 8089, CY Cergy Paris Universit\'e, 95302 Cergy-Pontoise Cedex, France}
\author{Mario Collura}
\affiliation{International School for Advanced Studies (SISSA), via Bonomea 265, 34136 Trieste, Italy}
\affiliation{INFN Sezione di Trieste, 34136 Trieste, Italy}

\begin{abstract}
We propose an enhanced Time-Dependent Variational Principle (TDVP) algorithm for Matrix Product States (MPS) that integrates Clifford disentangling techniques to efficiently manage entanglement growth. By leveraging the Clifford group, which maps Pauli strings to other Pauli strings while maintaining low computational complexity, we introduce a Clifford dressed single-site 1-TDVP scheme. During the TDVP integration, we apply a global Clifford transformation as needed to reduce entanglement by iteratively sweeping over two-qubit Clifford unitaries that connect neighboring sites in a checkerboard pattern.
We validate the new algorithm numerically using various quantum many-body models, including both integrable and non-integrable systems. Our results demonstrate that the Clifford dressed TDVP significantly improves entanglement management and computational efficiency, achieving higher accuracy, extended simulation times, and enhanced precision in computed observables compared to standard TDVP approaches. Additionally, we propose incorporating Clifford gates directly within the two-site 2-TDVP scheme. 
\end{abstract}

\maketitle

\paragraph{Introduction. ---}\label{s:introduction}
Simulating non-equilibrium dynamics of closed quantum many-body systems is a fundamental challenge in modern physics, whose difficulty primarily arises from the exponential size of the underlying Hilbert space and from the complex quantum correlations involved. The dynamics of these systems, where entanglement is generally expected to grow linearly in time (following a time-dependent volume law scaling of the entanglement entropy), demand powerful numerical methods to be addressed~\cite{Feynman1982,Daley_2022,xu2023herculean,Pashayan2020fromestimationof,anshu2023survey}. Matrix Product States (MPS) provide a robust numerical framework for simulating one-dimensional (1D) quantum many-body systems, offering a compact representation of 1D quantum states~\cite{Vidal_2004,Schollwock_2011,Paeckel_2019,Silvi_2019}. 

A standard approach for simulating the non-equilibrium dynamics of an MPS wave function involves the Trotterization of the evolution operator, thus transforming the Hamiltonian dynamics into a discrete quantum circuit of unitary gates acting locally between neighboring lattice sites (qubits) in the system. This method results in the well-known Time-Evolving Block Decimation (TEBD) scheme~\cite{Vidal_2004}, which efficiently handles real-time evolution by breaking down the global evolution into a sequence of local unitary gates.

Unlike TEBD, the Time-Dependent Variational Principle (TDVP) method evolves the MPS parameters directly within the MPS variational manifold~\cite{TDVP_2011,TDVP_2016}. TDVP ensures that the quantum state remains an optimal approximation within the MPS ansatz throughout the evolution, providing a more accurate representation of the system's dynamics, especially for long-time simulations and systems with strong entanglement growth. This approach is crucial for studying phenomena such as quantum quenches~\cite{Nishan_2022}, transport properties~\cite{PhysRevLett.117.207201}, and the growth of entanglement~\cite{Alba2018SciPost, Alba2016EntanglementAT}, where traditional methods would struggle with the exponentially growing Hilbert space.

Unfortunately, regardless of the chosen approach, a quantum system in a non-equilibrium state experiences a scrambling of quantum information, manifested as entanglement spreading throughout the system~\cite{Calabrese_2005,RevModPhys.80.517}. This process inevitably increases the complexity of the state. In the MPS representation, this results in the unbounded growth of the bond dimension. Consequently, beyond a certain threshold time, this exponential growth leads to a loss of accuracy in our approximate representation of the wave function, as the MPS can no longer efficiently capture the system's entanglement.

However, if we can manipulate the wave function to strategically reduce some of the quantum correlations and keep the entanglement level under control, this would be a promising strategy to extend the capabilities of current classical algorithms for simulating the unitary dynamics of quantum systems~\cite{PAECKEL2019167998}. 
Indeed, a remarkable insight comes from the field of quantum computing and information: circuits composed exclusively of Clifford gates (Hadamard, S and Controlled-NOT gates) can be efficiently simulated on a classical computer. 

Quantum states generated by Cliffords, known as stabilizer states, can exhibit volume law entanglement yet remain classically tractable (Gottesman Knill theorem)~\cite{Gottesman_1997, Gottesman_1998_1, Gottesman_1998_2}. The algorithm allowing this is known as tableau algorithm, as it consists in storing and evolving efficiently a compact table representing the Pauli strings that stabilize the state~\cite{Dehaene_2003, Aaronson_2004}. This facts highlight that while entanglement is a fundamental quantum resource, its presence alone does not necessarily make a computational problem intractable for classical devices.

Beyond entanglement, another essential quantum resource is `non-stabilizerness' or `magic', which significantly influences the complexity of quantum problems~\cite{gu2024magicinduced}. Similar to entanglement, non-stabilizerness has been rigorously defined within the paradigm of quantum resource theories, with various measures proposed to quantify it~\cite{Gross_2021,bu2023stabilizertestingmagicentropy,PhysRevB.109.174207,PhysRevA.99.062337,grewal2024improvedstabilizerestimationbell,Haug_2023_1,Haug_2023,Leone_2022,Leone_2024}. These measures have become instrumental in understanding the computational power of quantum systems, emphasizing that non-stabilizerness is a key factor in determining the classical hardness of simulating quantum systems~\cite{niroula2024phase,Lami_2023,mello2024retrieving,PRXQuantum.4.040317, Lami_2023,Heinrich_2019,fux2023entanglementmagic,tarabunga2024nonstabilizerness,PhysRevA.109.L040401,Haug_2023_2,frau2024nonstabilizerness}.

Successfully integrating MPS with stabilizer formalism~\cite{Aaronson_2004, https://doi.org/10.48550/arxiv.1711.07848,https://doi.org/10.48550/arxiv.0807.2876} poses significant challenges. Recently, initial progress in this direction has been made with the introduction of an efficient algorithm for finding the stabilizer Pauli strings of an MPS~\cite{tarabunga2024nonstabilizerness, Lami_2024_1}, the stabilizer tensor network ansatz~\cite{masotllima2024stabilizer}, and subsequently the Clifford enhanced MPS ($\mathcal{C}$MPS)~\cite{Lami_2024_2}. The latter are states obtained by applying a Clifford unitary to an MPS. 
An optimization algorithm for these states has been proposed in Ref.~\cite{qian2024augmenting} in case of ground state search, effectively augmenting the Density Matrix Renormalization Group (DMRG) with Clifford circuits. Hybrid Clifford tensor networks algorithms have been introduced also in Refs.~\cite{mello2024hybrid} and \cite{paviglianiti2024estimating}.

In this letter, we advance the research in this direction by exploring a novel approach that integrates a Clifford-based disentanglement scheme with the TDVP algorithm to reduce the complexity of simulating out-of-equilibrium quantum systems. \\

\paragraph{Preliminaries. ---}\label{s:preliminary}
We examine a system of $N$ qubits,
although our findings
can be easily extended to qudits. We designate the local computational basis $\{\ket{0}, \ket{1}\}$ as the eigenstates of Pauli matrix $\hat{\sigma}^3$, such that $\hat{\sigma}^3 \ket{s} = (-1)^s \ket{s}$. A general Pauli string can be formed as $\hat{\Sigma}^{\boldsymbol{\mu}} = \hat{\sigma}_{1}^{\mu_1} \hat{\sigma}_{2}^{\mu_2} \cdots \hat{\sigma}_{N}^{\mu_N} \in \mathcal{P}_{N}$, where the boldface superscript $\boldsymbol{\mu}$ represents the set $\{\mu_1, \dots, \mu_N\}$, with $\mu_j\in\{0,1,2,3\}$ and zero stays for the local identity operator. In the following, for simplicity, subscripts indicating the qubits will be omitted whenever they are not required. Pauli strings form a complete basis for any operator $\hat{O}$ acting on the many-body Hilbert space $\mathcal{H} = \{\ket{0}, \ket{1}\}^{\otimes N}$. Specifically, we can express $\hat{O}$ as $\hat{O} = \sum_{\boldsymbol{\mu}} O_{\boldsymbol{\mu}} \hat{\Sigma}^{\boldsymbol{\mu}}$, where the coefficients are determined by $O_{\boldsymbol{\mu}} = {\rm Tr} ( \hat{O} \hat{\Sigma}^{\boldsymbol{\mu}})/2^N$. This uses the orthogonality condition ${\rm Tr}(\hat{\Sigma}^{\boldsymbol{\mu}}\hat{\Sigma}^{\boldsymbol{\nu}}) = 2^N\delta_{\boldsymbol{\mu}\boldsymbol{\nu}}$.

When an operator undergoes a generic unitary transformation, i.e. $\hat{U} \hat{O} \hat{U}^{\dag}$, its complexity in the Pauli basis representation can grow arbitrarily. However, when this transformation is chosen from the set of Clifford unitaries $\mathcal{C}_N$, the complexity of that operator remains bounded.
Indeed, the Clifford group $\mathcal{C}_N$ on $N$ qubits, consists of unitary operators that map Pauli strings into Pauli strings under conjugation. Specifically, for any Clifford unitary $\hat{C} \in \mathcal{C}_N$ and any Pauli operator $\hat{\Sigma}^{\boldsymbol{\mu}} \in \mathcal{P}_N$, we have $\hat{C} \hat{\Sigma}^{\boldsymbol{\mu}} \hat{C}^\dag = \pm\hat{\Sigma}^{\boldsymbol{\nu}}$. Clifford transformations, which are generated by the Hadamard gate, phase gate, and CNOT gate, are fundamental to fault-tolerant quantum computation and quantum error correction \cite{Nielsen_chuang_2010,Gottesman_1997,Gottesman_1998_1,Gottesman_1998_2, Bravyi_2005}.

Let us stress that the CNOT gate, which is the only responsible for the entanglement generation, is a member of this group of operators. In practice, any unitary matrix $\hat{C} \in \mathcal{C}_N$ can produce significant entanglement when applied to a many-body state, yet it preserves the intrinsically low complexity associated with Pauli string operators. In turn, the operator entanglement of a typical local Hamiltonian will be bounded by $\sim\log(N)$. \\

\begin{figure*}[t!]
\includegraphics[width=\linewidth]{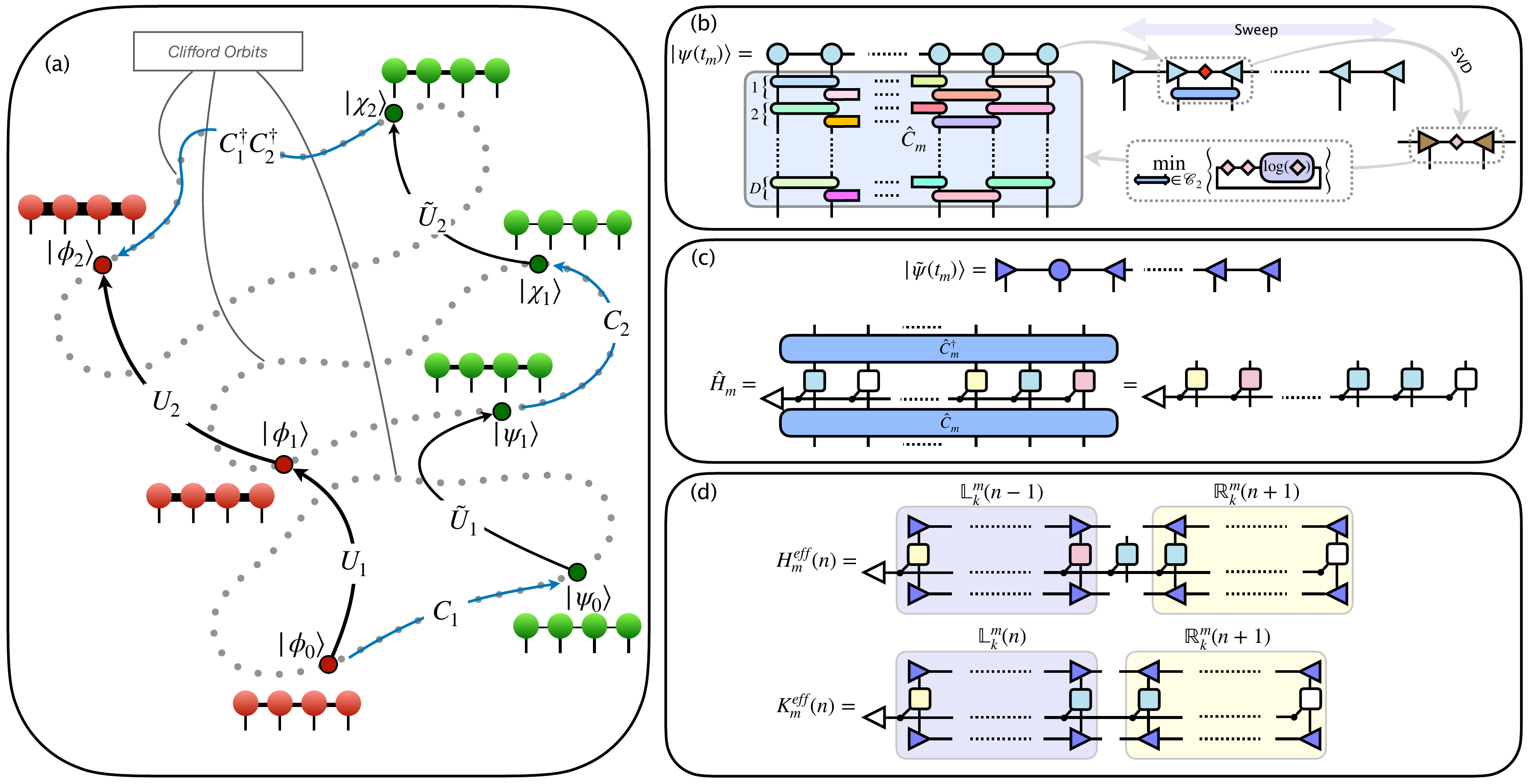}
\caption{Clifford dressed evolution. (a) The evolution of the state $\ket{\phi_0}$, driven by the successive application of $U_1$ and $U_2$, is transformed according to a modified evolution path $\tilde{U}_2 \tilde{U}_1$. In this new trajectory, each intermediate state is strategically adjusted within its respective Clifford orbit to significantly reduce the entanglement in its MPS representation. This approach ensures that the MPS wave-function maintains a lower entanglement throughout the evolution, enhancing the efficiency and accuracy of the simulation. (b) Clifford dressed TDVP steps, the disentangling routine optimally selects local Clifford two-qubit gates by minimizing the bipartite entanglement entropy, then sweeps along the chain up to a number of Clifford bi-layers of depth $D$. 
(c) At the end of the disentangling procedure, we obtain the Clifford dressed state $|\tilde{\psi}(t_m)\rangle$. The optimal Clifford transformation $\hat{C}_m$ is encoded using the stabilizer tableau and applied to the diagonal MPO Hamiltonian $\hat{H}_{m-1}$, preserving its diagonal structure. 
(d) The effective operators for the next TDVP step are iteratively constructed from the MPS of the Clifford dressed state in mixed canonical form, and the MPO of the Clifford dressed Hamiltonian. 
}
\label{fig:sketch}
\end{figure*}

\paragraph{Clifford disentangled 1-TDVP. ---}\label{s:TDVP}
The typical scenario we consider involves the dynamics generated by the Hamiltonian $H_0$ after preparing the system in a short-range correlated initial state (e.g., a product state) at time $t=0$. The subscript zero indicates that the Hamiltonian is initially in its ``bare" form, meaning it has not yet been modified by Clifford disentanglers.

We are generally interested in evolving the state and monitoring local observables which are experimentally relevant. This includes computing the expectation value of Pauli strings
$\bra{\psi(0)} e^{i \hat{H}_0 t} \hat{\Sigma}^{\boldsymbol{\mu}} e^{-i \hat{H}_0 t} \ket{\psi(0)}$. Typically, we track the time evolution of the state by discretizing time into small intervals $dt$. The state, approximated as an MPS with a fixed bond dimension $\chi$, is evolved using the single-site 1-TDVP scheme~\cite{TDVP_2016}.

However, the entanglement entropy typically grows unboundedly, rendering the MPS representation an inadequate wave-function ansatz beyond a relatively short time threshold. This is where Clifford-inspired disentangling strategies come into play.

Basically, at each step $m \in \{1, \dots\}$, corresponding to time $t_m = m \, dt$, we can apply a suitable Clifford transformation $\hat{C}_m$ to the state $\ket{\psi(t_m)}$ to reduce entanglement of the new state $|\tilde\psi(t_m)\rangle =\hat{C}_m \ket{\psi(t_m)}$. 

This disentangling routine, or ``entanglement cooling"~\cite{True_2022, Lami_2024_2, Odavic_2023, Chamon_2014, Shaffer_2014}, applied to the states induces a corresponding transformation of the Hamiltonian: $\hat{H}_m = \hat{C}_m \hat{H}_{m-1} \hat{C}_m^{\dagger}$. The latter can be efficiently performed using the stabilizer tableau formalism~\cite{Stim_2021}.

In practice, we can rewrite the original discrete evolution $\prod_{m=0}^{\lfloor t/dt \rfloor-1} e^{-i \hat{H}_0 dt} \ket{\psi(0)}$ as a \textbf{Clifford dressed} evolution (see Fig.~\ref{fig:sketch}(a)):
\begin{equation}\label{eq:clifford_dressed}
\prod_{m=0}^{\lfloor t/dt \rfloor-1} e^{-i \hat{H}_m dt}  \hat{C}_m\ket{\psi(0)},
\end{equation}
where $\hat{C}_0 = \hat{I}$ is the identity operator, and the product is understood to be in reverse order. Eq.~\eqref{eq:clifford_dressed} iteratively constructs the final $\mathcal{C}$MPS which results in $\hat{C}_{\lfloor t/dt \rfloor-1}\cdots\hat{C}_1\, e^{-i \hat{H}_0 t}\ket{\psi(0)}$~\cite{Lami_2024_2}. 
Additionally, at each time step $t_m$, any Pauli string we are interested in measuring must be transformed as $\hat{C}_{m} \cdots \hat{C}_{1} \hat{\Sigma}^{\boldsymbol{\mu}} \hat{C}_{1}^{\dagger} \cdots \hat{C}_{m}^{\dagger}$. 

Note that we are assuming the application of a Clifford disentangler at each time step of the TDVP integrator. However, this is not strictly necessary. We can choose to apply the entanglement cooling routine less frequently, for instance, every $k$ time steps, thereby reducing the computational overhead while still mitigating the growth of entanglement.

The disentangling routine constructs the optimal Clifford operator iteratively by sweeping over two-qubit Clifford unitaries that connect neighboring sites of the chain in a checkerboard pattern~\cite{True_2022,Lami_2024_2}. The number of bi-layers is denoted by $D$ (see Fig.~\ref{fig:sketch}(b)). The optimal Clifford two-qubit gate is selected by searching through the subset of 720 Clifford tableaux with positive sign~\cite{Stim_2021} and randomly choosing one of those that minimizes the von Neumann entanglement entropy. The sign of the tableau does not affect the local Singular Value Decomposition (SVD) of the MPS tensors. However, since the sign choice may influence subsequent minimizations, we reintroduce a random sign configuration from the $2^4$ possible options.
The procedure is repeated sequentially, sweeping back and forth over the entire chain.
 
We implement the 1-TDVP scheme because it is a symplectic integrator. Specifically, to evolve the state from $t_m$ to $t_{m+1}$, we project the dressed Hamiltonian $\hat{H}_m$ using the MPS tensors of the Clifford enhanced state 
$
|\tilde{\psi}(t_{m})\rangle = \mathbb{A}^{s_1}_{L} \cdots \mathbb{A}^{s_{n-1}}_{L} \mathbb{A}^{s_n}_{C} \mathbb{A}^{s_{n+1}}_{R} \cdots \mathbb{A}^{s_N}_{R} \ket{s_1, \dots, s_N},
$
which is in the mixed canonical form with respect to the central site $n$. This allows us to define the effective Hamiltonian (see Fig.~\ref{fig:sketch}(c))
\begin{equation}
H^{\text{eff}}_{m}(n) = \sum_{k}^{O(N)} J_k \, \mathbb{L}^{m}_{k}(n-1) \otimes \hat{\sigma}_{n}^{\mu^{m}_{k}} \otimes \mathbb{R}^{m}_{k}(n+1),
\end{equation}
which remains diagonal in the operator auxiliary dimension, as we keep $\hat{H}_m$ diagonal in the Pauli basis. Here, $J_k$ are the couplings associated with each Pauli string. A similar transformation is applied for $\hat{K}^{\text{eff}}_{m}(n)$ (we refer to Ref.~\cite{TDVP_2016} for more details on the TDVP algorithm in the context of MPS). \\

\begin{figure}[t]
    \centering
    \includegraphics[width=\linewidth]{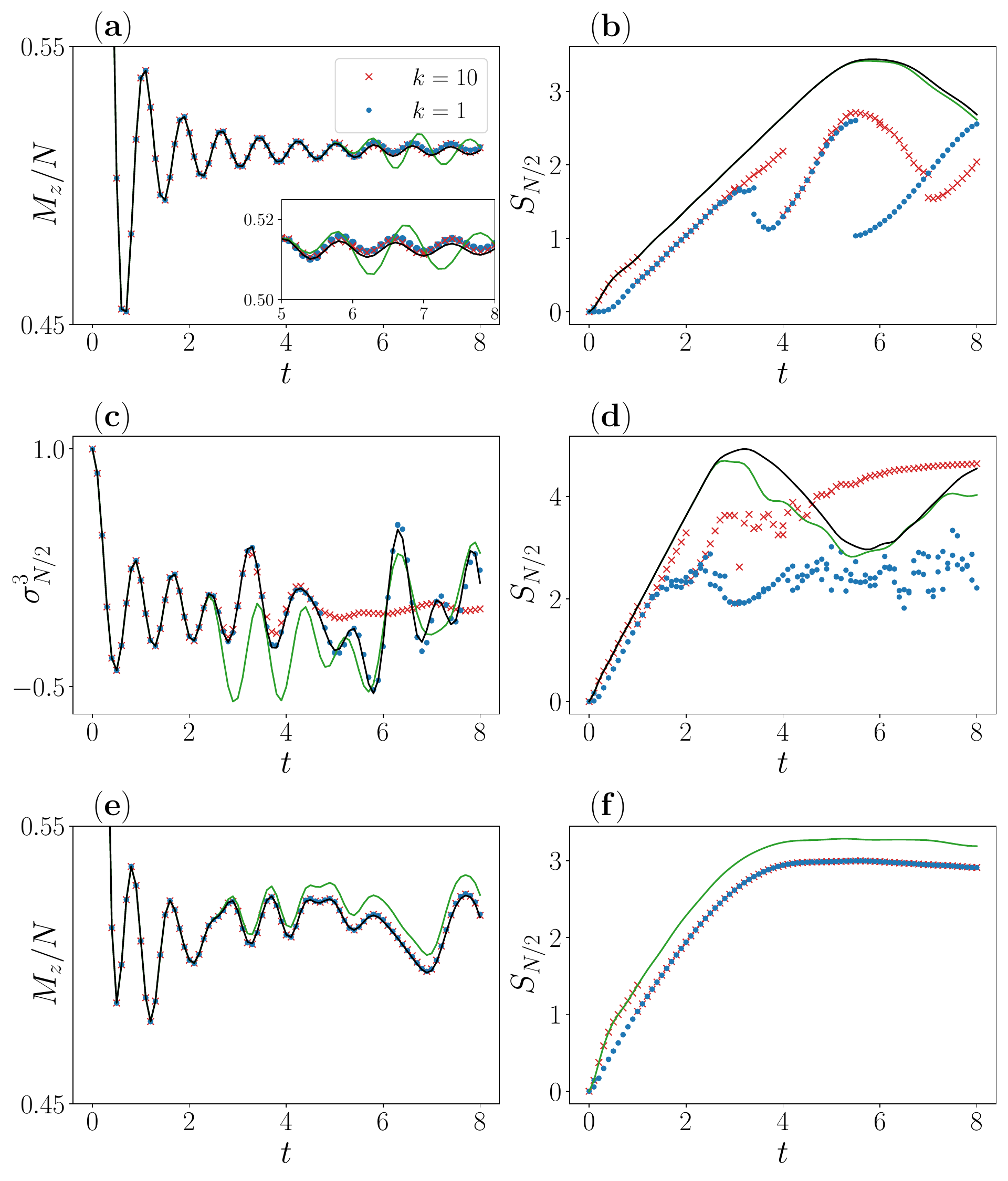}
    \caption{(a) Average magnetization and (b) half-chain entanglement entropy for the critical transverse field Ising model. (c) Half-chain magnetization and (d) half-chain entanglement entropy for the XX model. (e) Average magnetization and (f) half-chain entanglement entropy for the next-to-nearest-neighbors Ising chain. Black lines represent free-fermions (upper and central panels) or ED results (lower panel). Solid green lines are TDVP data. Markers indicate the Clifford disentangler, applied every $k$ time steps (see legend in the upper panel). In all panels we set $N=20,\ \chi = 128$.}
    \label{fig:mag_all_models}
\end{figure}

\begin{figure}[t]
    \centering
    \includegraphics[width=\linewidth]{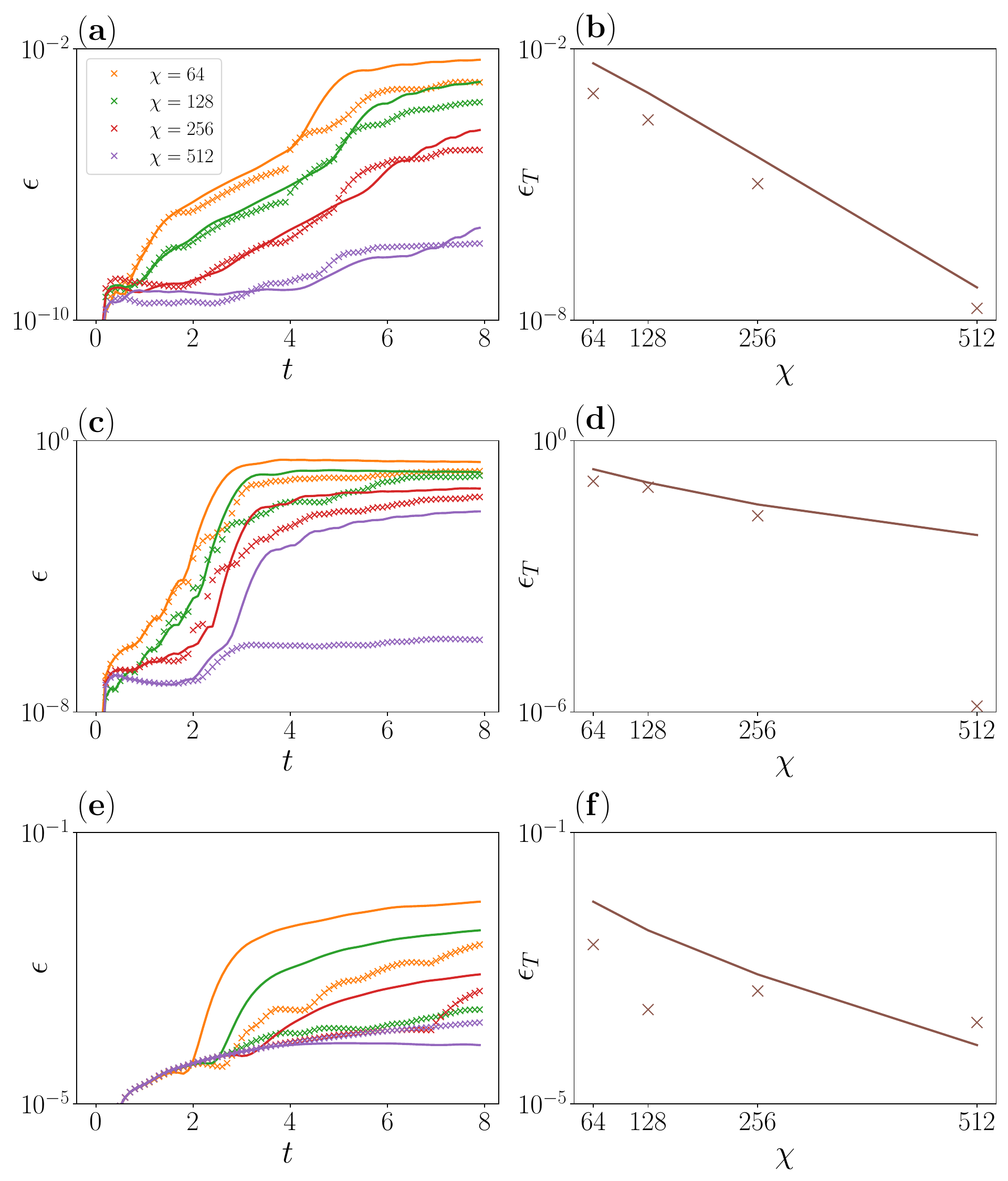}
    \caption{Error on expectation value of observables (same observables as in Fig.\ref{fig:mag_all_models}). (a)(c)(e) Integrated error $\epsilon (t)$, and (b)(d)(f) integrated error at final time $\epsilon_T$ ($T=8$). (a)(b) Critical transverse field Ising model. (c)(d) XX model. (e)(f) Next-to-nearest-neighbors Ising chain. Solid lines are TDVP data. Markers indicate the Clifford disentangler, applied every $k=10$ time steps. We set $N=20$ and explore different bond dimensions $\chi$ (see legend in the upper panel).}
    \label{fig:err_all_models}
\end{figure}

\paragraph{Numerical experiments. ---}\label{s:numerics}
We benchmark our algorithm on the following bare Hamiltonian
\begin{align}
    \hat{H}_{0}  =  &J^x_1\sum_{j=0}^{N-2} \hat{\sigma}^{1}_{j}\hat{\sigma}^{1}_{j+1} + J^y_1\sum_{j=0}^{N-2}\sigma^2_j\sigma^2_{j+1}+\notag\\ +&J^x_2\sum_{j=0}^{N-3}\sigma^1_j\sigma^1_{j+2} +h \sum_{j=0}^{N-1}\hat{\sigma}^{3}_{j}. \label{eq:Hbare}
\end{align}
Specifically, after fixing the MPS bond dimension $\chi$, that is the amount of employable resources, we compare the standard 1-TDVP with our novel strategy where the Clifford disentangler routine is invoked every $k$ time steps (for various values of $k$).

Firstly, we consider two integrable cases by setting $J_2^x=0$ in Eq.~\eqref{eq:Hbare}. Specifically, the critical quantum Ising chain $J^y_1=0$, $J^x_1=-h=1$  prepared in the fully polarized state $\ket{00 \dots 0}$ along the $\hat{z}$ direction, and, the XX model with $J^x_1=J^y_1=1$ and $h=0$ prepared in the Néel state $\ket{0101 \dots 01}$. Finally, we consider the non-integrable next-to-nearest-neighbors Ising model by setting $J_2^x=J_1^x=-h=1$ and $J^y_1=0$ prepared in $\ket{00 \dots 0}$.


During the time evolution, we observe the dynamics of the bipartite entanglement entropy at the midpoint of the chain, which shows discontinuities each time the Clifford disentangler is applied.

To evaluate the algorithm's effectiveness in reproducing the expectation values of local observables (which may become non-local after the Clifford dressing), we measure the half-chain magnetization $\hat{\sigma}^3_{N/2}$ for the XX model. Notice that this model admits a $U(1)$ symmetry, therefore, the total magnetization is conserved. Instead, for the Ising model, in both cases we measure the average magnetization $\hat{M}_z / N = \sum_{j=0}^{N-1} \hat{\sigma}^{3}_{j} / N$. These quantities, which are not conserved, experience a nontrivial evolution. 
The XX and critical nearest-neighbors Ising models are exactly solvable, allowing us to compare our numerical findings with the exact solutions obtained via free-fermion techniques. For the next-to-nearest-neighbors Ising model, instead, we resort to comparison with exact diagonalization (ED) calculations. 

We define a measure of the error up to time $t$ as the integrated distance between the exact expectation value of an observable $O(t)$ and the corresponding value obtained with either TDVP or Clifford enhanced TDVP denoted as $\tilde{O}_\chi(t)$ (for fixed $\chi$), i.e.\
\begin{equation}
    \epsilon(t) = \frac{1}{t} \int_0^t \abs{\tilde{O}_\chi(t') - O(t') } \, dt' \, .
\end{equation} 
We also define the integrated error at final time as $\epsilon_T = \epsilon(T)$. 
As mentioned, in our case $O$ corresponds to either the average magnetization or the on site half-chain magnetization.

Fig.~\ref{fig:mag_all_models} shows the results for the observable evolution and the half-chain entanglement entropy in all analyzed models for fixed bond dimension $\chi = 128$ and system size $N=20$. The corresponding panels in Fig.~\ref{fig:err_all_models} show the evolution of the error $\epsilon(t)$ for $k=10$.

For the critical Ising chain in Figs.~\ref{fig:mag_all_models}(a) and \ref{fig:mag_all_models}(b), Our approach limits the growth of entanglement entropy and enables us to achieve higher precision in magnetization dynamics compared to the standard 1-TDVP, which becomes ineffective around $t \approx 5$ due to the entanglement exceeding the bond dimension $\chi$. 
The most remarkable results are attained for the XX model, Figs.~\ref{fig:mag_all_models}(c) and \ref{fig:mag_all_models}(d), where at around $t \approx 2$ the 1-TDVP simulation breaks down. 
Clifford dressed evolution closely matches the exact solution for significantly longer times, demonstrating remarkable accuracy. 
Notice how the improved evolution obtains several order of magnitudes in precision for the same amount of resources.
Finally, Figs.~\ref{fig:mag_all_models}(e) and \ref{fig:mag_all_models}(f) refer to the non-integrable setup, where disentangling the evolved state becomes more challenging than in the previous scenarios. Nevertheless, the evolution of the Clifford dressed observable seems to match the one provided by ED for $\chi \leq 128$. The integrated error, contrarily to the other case studies, does not allow to claim a significant advantage. \\

\paragraph{Conclusion and Outlook. ---}\label{s:conclusion}

In this work, we introduce a novel algorithm to enhance the simulation of quantum time-evolution. We improve the standard single-site TDVP algorithm for MPS by incorporating optimized Clifford disentanglers in order to mitigate the growth of entanglement. The algorithm can be considered as an effective way to perform time evolution for the recently introduced states known as Clifford enhanced MPS~\cite{Lami_2024_2}. Our numerical experiments demonstrate the effectiveness of the method in managing controlled entanglement growth while accurately reproducing the expectation values of physical observables.

Several possible future research are apparent. First, improving the way Clifford disentanglers are optimized could yield further improvements. Various possibilities can be explored, such as introducing finite temperature in a Metropolis-like algorithm, or simultaneously optimizing multiple local Clifford gates. These approaches could help in avoiding sub-optimal local minima. Secondly, akin to the approach taken in Ref.~\cite{qian2024augmenting} for DMRG, a potential extension could leverage the two-site TDVP (2-TDVP). In this scenario, the two-qubit Clifford disentangler would be applied to disentangle the state just before performing the SVD. Once the optimal disentangler is identified, the effective Hamiltonian is mapped under conjugation, affecting the Pauli strings only locally. This approach would maintain locality while efficiently managing entanglement, albeit at the cost of sacrificing the symplectic nature of the integrator.

A fundamental theoretical question that remains unresolved is whether it is feasible to quantify the effectiveness of the ansatz used, thereby distinguishing between quantum states that can be successfully simulated using our method (i.e.\ those that can be significantly Clifford disentangled) and those for which this is not possible. This would be crucial to gain a deeper understanding of the benefits and limitations method. A possibility is a connection with the entanglement spectrum statistics, which should exhibit signatures of entanglement irreversibility~\cite{Chamon_2014, Shaffer_2014}. \\

\paragraph{Acknowledgments. --}
We acknowledge the use of Stim~\cite{Stim_2021} for the stabilizer formalism operations, ITensor for the tensor networks simulations~\cite{itensor} and QuTiP for the ED simulations~\cite{qutip}. 
We are particularly grateful to Martina Frau, Alessio Paviglianiti, Poetri Tarabunga, Emanuele Tirrito, Tobias Haug, Gerald Fux, Marcello Dalmonte and Rosario Fazio for inspiring discussions on the topic.
This work was supported by the ANR-22-CPJ1-0021-01, the ERC Starting Grant 101042293 (HEPIQ), the PNRR MUR project PE0000023-NQSTI, and the PRIN 2022 (2022R35ZBF) - PE2 - ``ManyQLowD''. \\ 


\bibliography{bib}

\end{document}